\pdfoutput=1

\documentclass[a4paper,fleqn]{cas-dc}

\usepackage[authoryear,longnamesfirst]{natbib}
\usepackage{pdflscape}
\usepackage{textcomp}

\def\tsc#1{\csdef{#1}{\textsc{\lowercase{#1}}\xspace}}
\tsc{WGM}
\tsc{QE}
\tsc{EP}
\tsc{PMS}
\tsc{BEC}
\tsc{DE}

\begin{document}
\let\WriteBookmarks\relax
\def\floatpagepagefraction{1}
\def\textpagefraction{.001}

\shorttitle{PATopics}

\shortauthors{Cecilio P  et~al.}

\title [mode = title]{PATopics: An automatic framework to extract useful information from pharmaceutical patents documents}                      



%



\credit{Conceptualization of this study, Methodology, Software}

\affiliation[1]{organization={Universidade Federal de São João del Rei (UFSJ)},
    country={Brazil}}

\affiliation[2]{organization={Universidade Federal de Minas Gerais (UFMG)},
    country={Brazil}}

\affiliation[3]{organization={Instituto de Investigação e Inovação em Saúde (I3S), Universidade do Porto},
    country={Portugal}}

\affiliation[4]{organization={Faculdade de Ciências Farmacêuticas de Ribeirão Preto, Universidade de São Paulo (USP)},
    country={Brazil}}

\author[1]{Pablo Cecilio}[style=chinese,orcid=https://orcid.org/0000-0002-4116-7021]
\fnmark[1]

\author[1]{Antônio Perreira}[style=chinese, orcid=https://orcid.org/0000-0001-5862-9272]


\credit{Data curation, Writing - Original draft preparation}

\author[3]{Juliana Santos Rosa Viegas}[style=chinese,orcid=https://orcid.org/0000-0002-8615-3758]

\author[2]{Washington Cunha}[style=chinese, orcid=https://orcid.org/0000-0002-1988-8412]

\author[2]{Felipe Viegas}[style=chinese, orcid=https://orcid.org/0000-0001-8121-8607]

\author[1]{Elisa Tuler}[style=chinese,orcid=https://orcid.org/0000-0003-3595-9978]
\author[4]{Fabiana Testa Moura de Carvalho Vicentini}[style=chinese, orcid=https://orcid.org/0000-0002-0842-9130]

\author[1]{Leonardo Rocha}[style=chinese,orcid=https://orcid.org/0000-0002-4913-4902]
\cormark[1]

\cortext[cor1]{Corresponding author}



\begin{abstract}
Pharmaceutical patents play an important role by protecting the innovation from copies but also drive researchers to innovate, create new products, and promote disruptive innovations focusing on collective health. The study of patent management usually refers to an exhaustive manual search. This happens, because patent documents are complex with a lot of details regarding the claims and methodology/results explanation of the invention. To mitigate the manual search, we proposed PATopics, a framework specially designed to extract relevant information for Pharmaceutical patents. PATopics is composed of four building blocks that extract textual information from the patents, build relevant topics that are capable of summarizing the patents, correlate these topics with useful patent characteristics and then, summarize the information in a friendly web interface to final users. The general contributions of PATopics are its ability to centralize patents and to manage patents into groups based on their similarities. We extensively analyzed the framework using 4,832 pharmaceutical patents concerning 809 molecules patented by 478 companies. In our analysis, we evaluate the use of the framework considering the demands of three user profiles -- researchers, chemists, and companies. We also designed four real-world use cases to evaluate the framework's applicability. Our analysis showed how practical and helpful PATopics are in the pharmaceutical scenario.
\end{abstract}




\maketitle

\section{Introduction}
The history of pharmaceutical patents is different from general product patents. In the pharmaceutical context, there is a great paradox between -- concern for public health and profits and income. In this scenario, patenting is not only related to the innovation but also to protecting the innovation from copies during a time that sufficiently covers the values applied in research and development. The consequent search for medicines that increasingly helps to cure or treat diseases/conditions and improves people's quality of life is what drives researchers to innovate, create new products, and promote disruptive innovations~\citep{zanotto2021stroke}, but it requires high investment and funding~\citep{Garattini2022, Khachigian2020, Malerba2015}.\looseness=-1

The average time from discovering a promising and effective molecule to reach the pharmacy shelf varies between ten and fifteen years~\citep{Berdigaliyev2020}. Millions of dollars of investment from basic research to preclinical and clinical trials are mandatory to launch a product. The manufacturing of a pharmaceutical product is a process that takes time. During the development stages, there is a chance that the product will be discontinued or prolonged when the cost-benefit to health is no longer favorable to this product. This may occur due to several factors, such as toxicological features, metabolization, and adverse effects. In this way, thousands of dollars are circulating between one product and another, one innovation and another, which are discouraged along the way~\citep{Wouters2020}.\looseness=-1

In fact, patenting ensures the protection of intellectual property rights and guarantees that the high investments return to the investors. Nowadays, the time granted to pharmaceutical patents is about ten years, which can be extended to fifteen or twenty years, in proportion to the expenditure on Research and Development (R\&D)~\citep{cecilio2023framework}. Patent time is also related to the product's market value, the number of potential users and claims~\citep{Hwang2021}.\looseness=-1

Studies indicate that works developed in the academy, such as results of Ph.D. projects, are auspicious and more patentable than products developed in companies. Still, several small companies develop patentable products that they do not deposit in databases. Several countries do not have a database for patents and still, not all patents are deposited in these repositories. Thus, the lack of registration of patents' data in public repositories, in sum with decentralized data, makes the search difficult. Then, much information about intellectual properties is not found, and consequently, many products that could go through technology transfer processes are ``lost''~\citep{Genin2021, Holgersson2019, Orviska2019}.\looseness=-1

Regarding the study of patent management, to the best of our knowledge, professionals have exhaustive manual work to get information about patents. This happens because patent documents are complex, with a lot of information and details. Therefore, sometimes, professionals have to read an extensive document to obtain simple information or detail.\looseness=-1

On the other hand, Topic Modeling is an extensive research field in Computer Science that gained much attention from the scientific community in recent years. The main reason is its applicability in several social and practical contexts. Topic Modeling is the machine learning task that \textbf{automatically} extracts “implicit” topics from a collection of documents and assigns the most probable ones to each document~\citep{viegas2019,junior2022evaluating,luiz2018feature}.\looseness=-1

In this context, Topic Modeling can help create a single environment for collecting pharmaceutical patents and also create a relationship between them based on their main subject. Creating topics summarizing the patents will give insights regarding the patentability process, the molecules involved, the main subjects, the innovation, and updates. In addition, Topic Modeling in pharmaceutical patents scenario will favor the search for patents. These contributions would be inserted in several users' profiles, and we will be highlighted three of them, (i) the academics and employers who work with patents search, (ii) the chemists and patents developers, and (iii) companies and industries who use, buy or applied the transfer technology of patents.\looseness=-1

In this work, we propose \textbf{PATopics}, a Topic Modeling framework specially designed to automatically fetch pharmaceutical patents from the Web and create semantically topics using Topic Modeling approaches to summarize the patents. The framework has a summarization interface showing quantitative analysis related to the topics. The main research questions (RQ) are: \textbf{RQ1:} \textit{Does PATopics capable of summarizing Pharmaceutical Patents in coherent topics?} \textbf{RQ2:} \textit{Do Pharmaceutical topics carry relevant information to help professionals?} The goal of this work is to create a framework (a.k.a PATopics) capable of summarizing the information about the patents and providing helpful information for users without the need to manually seeking for them.\looseness=-1

\section{Materials and Methods}
\subsection{Framework construction}
In this Section, we revisit our framework description, formally presented in~\citep{Viegas2022}. In the context of the current proposal, our goal is to use the proposed framework~\citep{Viegas2022} to build semantic topics for pharmaceutical patents and use these topics to correlate them with inventors, chemical compounds, and pharmaceutical companies. We divided the framework instantiation into four main steps: (i) Data representation, (ii) Topic modeling decomposition, (iii) Correlation among entities, and (iv) Summary Interface.\looseness=-1

\subsubsection{Data representation}\label{sec:data_repr}
In this step, several data representation strategies can be used to represent the textual description of pharmaceutical patents. In~\citep{Viegas2022}, the framework has four types of data representation implemented: TF-IDF~\citep{tfidf,cunha2020extended}, TF-IDF with bigrams, CluWords~\citep{viegas2019}, CluWords with bigrams (data representation that combines the CluWords representation exploiting bigrams). Briefly speaking, the TF-IDF representation is one of the most traditional forms of representing textual data~\citep{cunha2022comparative,cunha2023effective}. It is a fixed-length vector where each index represents the word in the vocabulary collection. Bigrams are usually used to enrich the data representation, where the two adjacent words are included as a unique element in the vocabulary of the collection. We explored the \textit{gemsim} function Phrases to build the bigrams. To reduce the number of combinations, we ignore all bigrams with score $\left(word_a, word_b \right) < 0.5$, where the function score $\left( \cdot \right )$ returns the occurrence in the collection. The Cluwords representation incorporates semantic information to enrich the textual information. The method has three main steps: (a) Clustering -- it exploits the nearest neighbors' approach to capture the semantic relatedness; (b) Filtering -- it filters possible noises in the semantic neighborhood; (c) Weighting -- it combines the TF-IDF representation with the semantic neighborhood by a weighting scheme.\looseness=-1 

In previous works, considering scientific articles in Computer Science~\citep{Viegas2022} area, we evaluated the quality of the topics built considering the four data representations. Our results showed that the CluWords with bigrams outperformed the other data representations. Thus, in this work, we choose the CluWords with bigrams to represent the textual data of the pharmaceutical patents.\looseness=-1

\subsubsection{Topic modeling decomposition}\label{sec:topic_deco}
In this step, the framework exploits the topic modeling method called Non-negative Matrix Factorization (NMF)~\citep{MENG20181277,cunha2021cost,viegas2018semantically}. The NMF method is a matrix factorization where an input matrix A is decomposed into two matrices $H \in \mathbb{R}^{n \times k}$ and $W \in \mathbb{R}^{k \times m}$. The goal is to find the $k$-dimensional approximation that satisfies $A \approx H \times W$. Each k-dimension is represented as a topic in the NFM method. The matrix H codifies the relationship between documents and the topics (k-dimensions), whereas the matrix W codifies the relationship between the words and topics.\looseness=-1

\subsubsection{Correlation among entities}
In this case, we consider as input the data collection with a textual description of pharmaceutical patents and the matrices H and W decomposed by the NMF method. Following the example, consider that the patent $i^{th}$ of the matrix H deals mainly with the topic "Cancer treat." while the patent $j^{th}$ deals with "Autoimmune". For this example, each patent has one or more inventors, so it is possible to highlight which topics are most related to inventors through the relationships between patents and topics found. Likewise, as the inventors of a patent work for research companies, it is also possible to highlight companies by topics, considering the relationship between patents and topics. The strategy consists of manipulating the matrices provided by the NMF that correlate topics and patents, introducing information from inventors and their companies, as in the following example - considering the matrices $H$ and $W$ for three topics. First, each topic is identified by analyzing the matrix $H$ and discovering which words are most strongly associated with each topic. Assuming the example where the first topic is mainly associated with "Cancer treat.", the second with "Autoimmune" and the third with "Pain treat.". Analyzing the W matrix that relates documents and topics, taking as an example the first matrix in Table~\ref{tab:calc_topics}, which contains three patents, where each position presents the "relevance" of the theme for the document. Thus, grouping and summing the values of topics achieved for patents that belong to the same inventor leads us to the second matrix in Table\ref{tab:calc_topics}. Assuming that the three patents in the first matrix belong to the first inventor in the second matrix, inferring the "relevance" of each topic for this inventor.\looseness=-1

\begin{table}[h!]
\caption{Calculating Inventor Contributions to Topics.}
\begin{center}
	\begin{tabular}{cccc}
		\multicolumn{4}{c}{(a) Resulting NMF}\\
		\hline
		&Cancer treat. & Autoimmune & Pain treat.\\
		Patent 1 & 30 & 70 & 10 \\
		Patent 2 &20 & 65 & 40 \\
		Patent 3 &17 & 80 & 8
	\end{tabular}

	\Large{$ \Downarrow  $}
\end{center}
\begin{center}
	\begin{tabular}{cccc}
		\multicolumn{4}{c}{(b) Inventor's Pertinence by Topic}\\
		\hline
		&Cancer treat. & Autoimmune & Pain treat.\\
		Inventor 1 & 67 & 215 & 58 \\
		Inventor 2 & 47 & 150 & 18 \\
		Inventor 3 & 20 & 65 & 40
	\end{tabular}
	
	\Large{$ \Downarrow  $}
\end{center}
\begin{center}
	\begin{tabular}{cccc}
		\multicolumn{4}{c}{(c) Normalized Inventor's Pertinence by Topic}\\
		\hline
		&Cancer treat. & Autoimmune & Pain treat.\\		
		Inventor 1 & 20\% & 63\% & 17\% \\
		Inventor 2 & 22\% & 70\% & 8\% \\
		Inventor 3 & 16\% & 52\% & 32\%
	\end{tabular}
	\end{center}
\vspace{0.3cm}
\label{tab:calc_topics}
\end{table}

Considering the matrix $W$ of the NMF method, the same process can be applied to all inventors. We can calculate the distribution among the topics to which each inventor's patents relate. By normalizing the lines representing the inventors in the second matrix of Table~\ref{tab:calc_topics}, it is possible to measure the impact of each inventor's research on each topic (Third matrix of Table~\ref{tab:calc_topics}). This process is repeated to extract how relevant the companies' patents are for each topic.\looseness=-1

\subsubsection{Summary interface}
The framework has an intuitive interface visualization that summarizes the topics, their associations with inventors/companies, and the main molecules involved. This way, all the analyses proposed in this Section can be carried out quickly and efficiently. The interface presents data related to topics and their correlations with patents, inventors/companies, and molecules through a large set of visual metaphors in a web application, which provides tools capable of presenting the same result in different ways, including graphs, heat maps, tables, and interactive search methods. This Web App is available on our website.\footnote{\url{https://labpi.ufsj.edu.br/patopics/}. \textbf{username}: user-test - \textbf{password}: avaliacao}

\subsection{Data collecting and cleaning}
In this Section, we describe the steps to instantiate the PATopic framework. To build the dataset of pharmaceutical patents, we extracted 4,832 patents from the WizMed platform (\url{https://wizmed.com/drug-patent-database}). We selected the patents in English and published them from 2003 to 2020. The dataset contains the following patent information: \textbf{1.} Patent Identifier; \textbf{2.} Title of the Patent; \textbf{3.} Description; \textbf{4.} Abstract; \textbf{5.} Drug; \textbf{6.} Company; \textbf{7.} URL; \textbf{8.} Strength; \textbf{9.} Trade Name.\looseness=-1

To build the data representation described in Section~\ref{sec:data_repr}, we exploited the Title and description fields of the patents. In terms of textual preprocessing, we performed stop-word removal (using the standard SMART list). We also removed words such as adverbs, verbs, and intensifiers. In terms of the topics built by the topic modeling method (Section~\ref{sec:topic_deco}), we computed three settings of topics summarization -- 10, 20, and 30. We also considered three settings -- 10, 20, and 30 -- of top words to represent the topics.\looseness=-1

\section{Results and discussion}
\subsection{Framework overview}
\begin{figure*}[ht]
	\centering
	\includegraphics[scale=0.4]{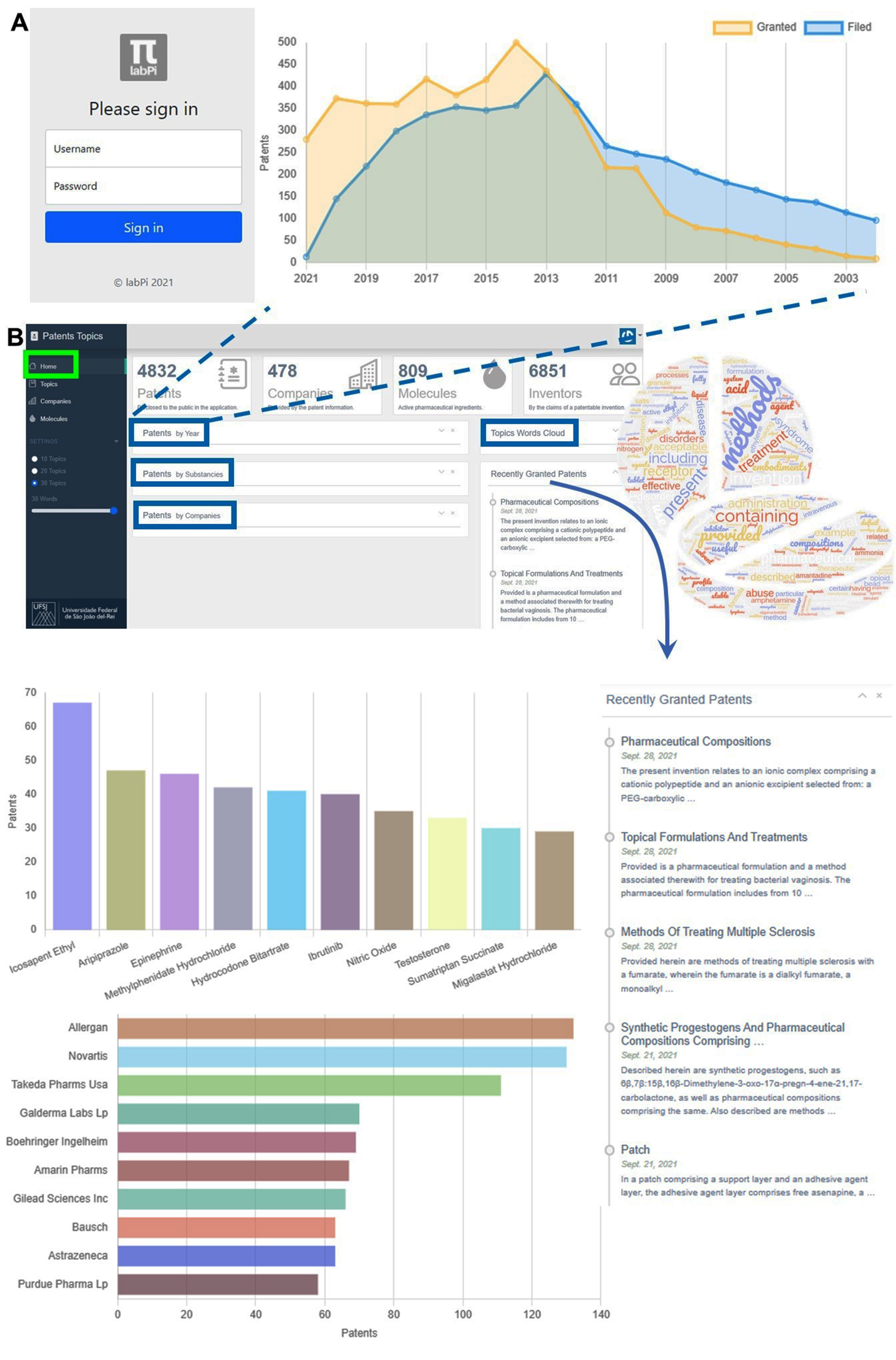}
	\caption{Framework interface (A), which users access by login, and after logging, the homepage (B) presents an easily navigable interface showing the number of patents collected, companies involved, and related molecules and their respective inventors. The homepage also exhibits graphs of patents per year, from 2003 to 2021; patents per molecule, and patents per company. In the right exhibit, the topics words cloud and the most recent patents are included.}
	\label{fig:1}
\end{figure*}

The framework allows the registration of users who can access their personal accounts by user login (Figure~\ref{fig:1}-A). The framework's interface is easily navigable and intuitive. On the homepage, we have access to the number of patents, companies, related molecules, and the number of inventors who claim a patentable invention. 4,832 patents were collected from the WizMed platform and refer to 809 molecules patented by 478 companies. These patents comprise 6,851 inventors in the period from 2003 to 2021 (Figure~\ref{fig:1}-B). Moreover, on the homepage, the graphs of patents by year (granted and filed) are displayed, as also the correlation of patents by molecules and companies. The most recent patents are highlighted on the panel, as well as the word cloud generated by the collected patents.\looseness=-1

\begin{figure*}[ht]
	\centering
	\includegraphics[scale=0.45]{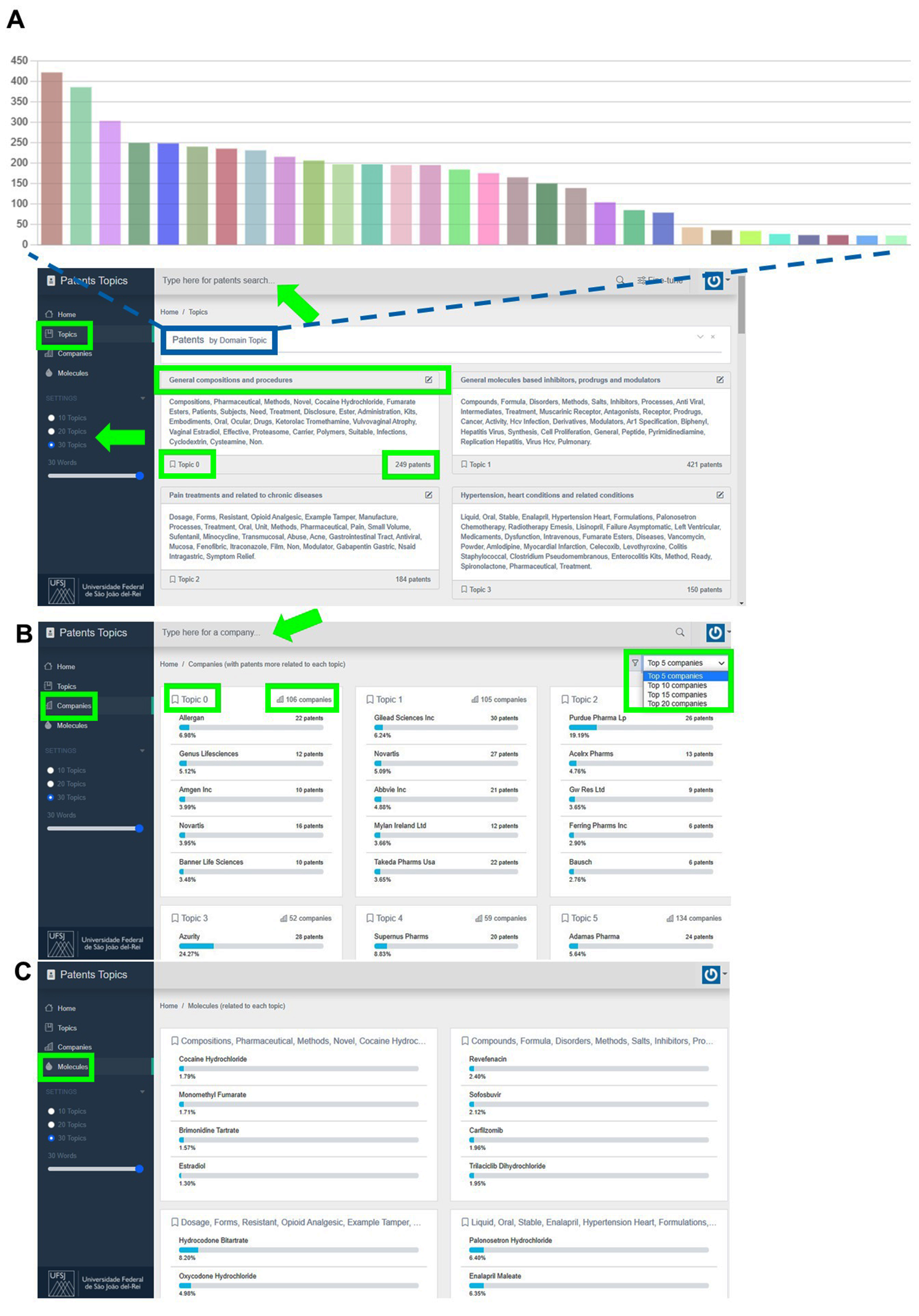}
	\caption{The Topics section (A) comprises a search bar and the generated topics by the word groups are described with the possibility of editable title and the number of patents per topic is visible; The companies' section (B) has a search bar (by the company) and they are distributed in the generated topics at 5, 10, 15 or 20 companies per topic; The molecules section (C) where the molecules mentioned per topic are highlighted.}
	\label{fig:2}
\end{figure*}

The Topics tab (Figure~\ref{fig:2}-A) is built according to the words, 30 topics in this present study, which can be changed to 10 or 20. The number of words can also be adjusted. In this case, we adopted 30 words, which can also be changed to 10 or 20 words. In this way, the words generate the topics and according to them, the topic has been titled by a person with knowledge of pharmaceutical issues and this title is easily editable. Therefore, the evaluation and results of the Topics built are not automatic and they should be handled by a professional in the field of the study.\looseness=-1

It is possible to click on the Topic, access the patents covered by it, and even access each patent individually. When entering a Topic, the number and title of the patents are immediately observed, just below are the data: patent date, company, molecule, concentration (if it exists), and a brief. Right below the data, there is a bar representing the Topics and the percentage of each Topic that the patent includes. The patent is selected by the topic in which it has the highest percentage of similarity. For instance, some patents will present 100\% for the Topic in which they are inserted, but others will present percentages distributed between 2 or more topics. Some topics have more patents than others. For example, Topic 1 has 421 patents, and Topics 18 and 28 have 23 patents each. This grouping occurs by the relationship between the words.\looseness=-1

The Companies tab (Figure~\ref{fig:2}-B) is built based on the generated topics. However, it can adjust the number of exhibited companies per topic to 5, 10, 15, or 20. It is possible to click on the Company and access the company data, which comprises the number of patents per Topic, the number and title of each patent and it is possible to click on every patent and access the details.\looseness=-1

The Molecules tab (Figure~\ref{fig:2}-C) is built based on the most patentable substances in each Topic. It is possible to observe the percentage of each molecule in each one and to access the molecule data, which comprises the related patents in every related Topic.\looseness=-1

In a nutshell, the framework and its sections allow quick access to patents and their respective details that it would only get by performing more complex and time-consuming searches. Therefore, creating an environment that includes all these data will bring agility to any user. In addition, it is possible to adjust the search and consequently adjust the topics according to the desired objective.\looseness=-1

\subsection{Topic analysis}
\begin{figure*}[ht]
	\centering
	\includegraphics[scale=0.4]{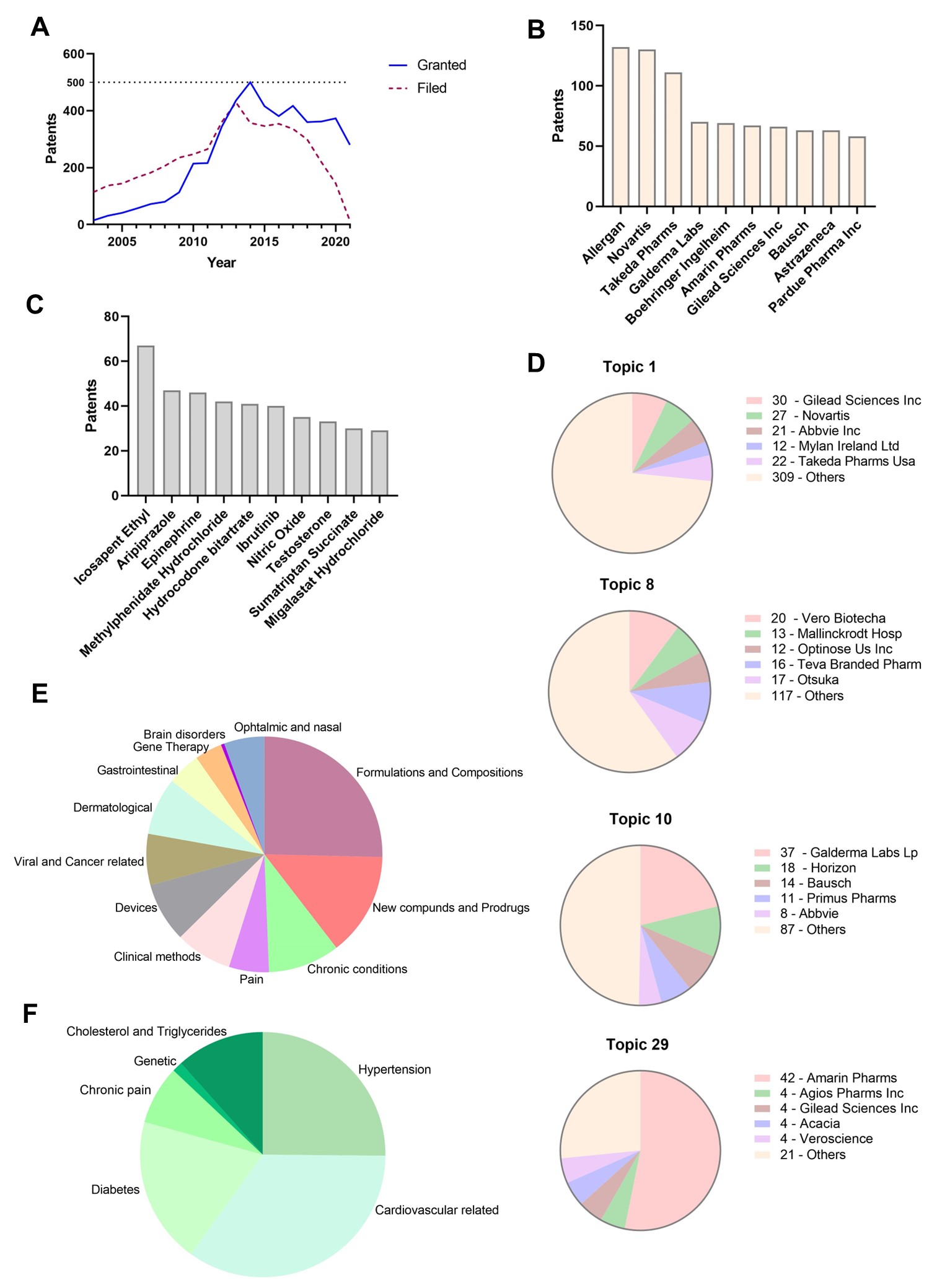}
	\caption{Summary of insights and results regarding PATopics. Quantitative analysis between patents and (A) years, (B) companies, (C) molecules, and regarding (D) Topics covered by companies (\%). (E) The main subjects involved in the collected pharmaceutical patents are Formulations and compositions, new compounds and prodrugs, chronic conditions, pain, clinical methods, devices, viral and cancer-related, dermatological, gastrointestinal, gene therapy, brain disorders, ophthalmic and nasal; showing with details the main patented (F) chronic conditions which are hypertension, cardiovascular-related, diabetes, chronic pain, genetic diseases and, cholesterol and triglycerides related.}
	\label{fig:3}
\end{figure*}

The PATopics allows the correlation between the patents by years. Figure~\ref{fig:3}-A shows that the number of patents has increased over the years, reaching a peak in 2014. Since 2014 the numbers have reduced, reaching a drop in the 2020-2021 biennium. The pandemic and, consequently, work reduction worldwide may be related to this drop in patent numbers.\looseness=-1

All output data are sectioned into Topics, Companies, and Molecules, and they summarize the input data. Table 2 shows the 30 Topics, their respective words, title, and the number of patents. Topic 1, entitled - General molecules-based inhibitors, prodrugs, and modulators, is the topic with the highest number of patents (421 patents), followed by Topic 5, entitled – Clinical methods (385 patents), and Topic 6, entitled - New compounds and prodrugs (303 patents).\looseness=-1

\begin{table*}[htb]
\centering
\scriptsize
\caption{Overview of the Topics (1-10) built by PATopics framework, its words, title, and the number of patents.}
\begin{tabular}{llll} \toprule
\textbf{Topic} & \textbf{Topic Title}  & \textbf{Words} & \textbf{Number of patents} \\ \midrule
0              & General compositions and procedures                                                                                                & \begin{tabular}[c]{@{}l@{}}Compositions, Pharmaceutical, Methods, Novel, Cocaine Hydrochloride, \\ Fumarate Esters, Patients, Subjects, Need, Treatment, Disclosure, Ester, \\ Administration, Kits, Embodiments, Oral, Ocular, Drugs, Ketorolac Tromethamine, \\ Vulvovaginal Atrophy, Vaginal Estradiol, Effective, Proteasome, Carrier, Polymers, \\ Suitable, Infections, Cyclodextrin, Cysteamine, non.\end{tabular}                                                                                               & 249                        \\ \hline
1              & \begin{tabular}[c]{@{}l@{}}General molecules-based inhibitors, \\ prodrugs and modulators\end{tabular}                             & \begin{tabular}[c]{@{}l@{}}Compounds, Formula, Disorders, Methods, Salts, Inhibitors, Processes, Anti-Viral, \\ Intermediates, Treatment, Muscarinic Receptor, Antagonists, Receptor, Prodrugs, \\ Cancer, Activity, Hcv Infection, Derivatives, Modulators, Ar1 Specification, \\ Biphenyl, Hepatitis Virus, Synthesis, Cell Proliferation, General, Peptide, \\ Pyrimidinediamine, Replication Hepatitis, Virus Hcv, Pulmonary.\end{tabular}                                                                          & 421                        \\ \hline
2              & \begin{tabular}[c]{@{}l@{}}Pain treatments and related to \\ chronic diseases\end{tabular}                                         & \begin{tabular}[c]{@{}l@{}}Dosage, Forms, Resistant, Opioid Analgesic, Example Tamper, Manufacture, \\ Processes, Treatment, Oral, Unit, Methods, Pharmaceutical, Pain, Small Volume, \\ Sufentanil, Minocycline, Transmucosal, Abuse, Acne, Gastrointestinal Tract, \\ Antiviral, Mucosa, Fenofibric, Itraconazole, Film, Non, Modulator, \\ Gabapentin Gastric, Nsaid Intragastric, Symptom Relief.\end{tabular}                                                                                                      & 184                        \\ \hline
3              & \begin{tabular}[c]{@{}l@{}}Hypertension, heart conditions \\ and related conditions\end{tabular}                                   & \begin{tabular}[c]{@{}l@{}}Liquid, Oral, Stable, Enalapril, Hypertension Heart, Formulations, \\ Palonosetron Chemotherapy, Radiotherapy Emesis, Lisinopril, \\ Failure Asymptomatic, Left Ventricular, Medicaments, Dysfunction, \\ Intravenous, Fumarate Esters, Diseases, Vancomycin, Powder, Amlodipine, \\ Myocardial Infarction, Celecoxib, Levothyroxine, Colitis Staphylococcal, \\ Clostridium Pseudomembranous, Enterocolitis Kits, Method, Ready, \\ Spironolactone, Pharmaceutical, Treatment.\end{tabular} & 150                        \\ \hline
4              & \begin{tabular}[c]{@{}l@{}}Formulations with controlled, \\ sustained or extended release \\ of drugs\end{tabular}                 & \begin{tabular}[c]{@{}l@{}}Release, Extended, Sustained, Oral, Profile, Component, Formulation, \\ Topiramate, Immediate, Dosage, Period, Tablet, Ion Exchange, Resin Complex, \\ Amantadine, Time, Hours Longer, Attention Deficit, Hyperactivity Disorder, \\ Preparations Oxcarbazepine, Lag Time, Methylphenidate, Active, Inventive, \\ Administration, Matrix, Optional Immediate, Mammalian Subjects, \\ Pathological, Polymer.\end{tabular}                                                                     & 215                        \\ \hline
5              & Clinical methods                                                                                                                   & \begin{tabular}[c]{@{}l@{}}Patient, Method, Dose, Salt, Administration, Amantadine, Disease, Treatment, \\ Need, Effective, Neurological, Period, Dravet Syndrome, Example, Condition, \\ Pirfenidone, Human, Excipients, Mifepristone, Parkinson Disease, Pharmaceutical, \\ Inhibitor, Formulation, Comprising, Androgen Cyclic, Cyclic Macrocyclic, \\ Macrocyclic Ketone, Male Hypogonadism, Thickening, Blood.\end{tabular}                                                                                        & 385                        \\ \hline
6              & New compounds and prodrugs                                                                                                         & \begin{tabular}[c]{@{}l@{}}Compound, Formula, Salt, Alkyl, Group, Solvate, Activity, Hydrogen, Cancer, \\ Treatment, Stereoisomer, Structure, Method, Derivatives, Human, \\ Receptor Antagonists, Peripheral, Pharmaceutical, General, Basic Addition, \\ Prodrugs Phosphoramidate, Modulating Kinases, Mammals, Viral Infections, \\ Inhibitor, Tetracycline, Heterocyclic, Atom, Nucleoside, Lower.\end{tabular}                                                                                                     & 303                        \\ \hline
7              & \begin{tabular}[c]{@{}l@{}}Methods and compositions for \\ genetic disorders\end{tabular}                                          & \begin{tabular}[c]{@{}l@{}}Fabry Disease, Galactosidase, Migalastat, Renal Impairment, Patients Hek, \\ Amenable Mutations, Methods, Treatment, Salt, Specific Pharmacological, \\ Chaperone, Vitro Vivo, Pharmacological Chaperone, Lysosomal Storage, \\ Line, Mutant, Response, Cell, Disorder, G9331a, Gla, Intron, Splice Site, \\ Sequence, Nucleic, Globotriaosylsphingosine, Nausea Vomiting, Ckd, \\ Diabetes Mellitus, Gene.\end{tabular}                                                                     & 36                         \\ \hline
8              & Delivery systems and devices                                                                                                       & \begin{tabular}[c]{@{}l@{}}Delivery, System, Nitric Oxide, Gas, Transdermal, Surface, Nasal, Vehicle, \\ Active, Nosepiece, Viscosity, Pulmonary, Unit, Administration, \\ Polyorthoester Polymer, Buprenorphine Sustained, Polar Aprotic, Agent, \\ Sustained, Conversion Nitrogen, Dioxide No2, Antioxidant Ascorbic, \\ Exemplary Gas, Solvent, Inhalation, Estrogen, Data, Method, Flowable, \\ Valve Assembly.\end{tabular}                                                                                        & 195                        \\ \hline
9              & \begin{tabular}[c]{@{}l@{}}Inhibitors related to viral and \\ cancerous process\end{tabular}                                       & \begin{tabular}[c]{@{}l@{}}Diseases, Inhibitors, Bruton Tyrosine, Conditions, Kinase, Treatment, \\ Btk Inhibitor, Autoimmune, Cancer Lymphoma, Inflammatory, Activity, \\ Agents, Protein, Amino Phenoxyphenyl, Pyrazolo Pyrimidin, \\ Subpopulation Lymphocytes, Expression, Methods, Pharmaceutical, \\ Therapeutic, Solvates, Derivatives, Indazole Derivatives, Janus, \\ Appearance Blood, Profile Biomarkers, Sufficient Increase, Specification, \\ Salts, Tyrosine Kinases.\end{tabular}                       & 231                        \\ \hline
10             & \begin{tabular}[c]{@{}l@{}}Dermatological conditions \\ and treatments\end{tabular}                                                & \begin{tabular}[c]{@{}l@{}}Topical, Diclofenac, Gel, Formulation, Skin, Application, Transdermal Flux, \\ Sodium Superior, Pain Osteoarthritis, Treatment, Properties, Benzoyl Peroxide, \\ Steroid, Effective, Compositions, Relief, Ivermectin, Disorders, Dermatological, \\ Acne, Suspension, Water, Users, Novel Preclinical, Clinical Information, Alcohol, \\ Psoriasis Dermatoses, Non-Foaming, Agent, Area.\end{tabular}                                                                                       & 175                        \\  \hline
11             & \begin{tabular}[c]{@{}l@{}}General formulations, methods and \\ new administration forms \\ and devices\end{tabular}               & \begin{tabular}[c]{@{}l@{}}Formulations, Methods, Administration, Treatment, Bendamustine, Pharmaceutical, \\ Application, Oral, Suitable, Intravenous, Bromocriptine Mesylate, \\ Sublingual Fentanyl, Glycemic Control, Tranexamic, Manufacturing, Disorders, \\ Spray Sublingual, Salt, Available, Diabetes, Doxepin, Embodiments, Injectable, \\ Sleep, Oil, Small Volume, Testosterone Deficiency, Fentanyl, Non, therewith.\end{tabular}                                                                          & 248                        \\ \hline
12             & \begin{tabular}[c]{@{}l@{}}Cardiovascular conditions \\ and related\end{tabular}                                                   & \begin{tabular}[c]{@{}l@{}}Cardiovascular Related, Inter, Fatty, Stable, Acid, Diseases, Pharmaceutical, \\ Methods, Compositions, Rhinitis, Tetracycline Antibiotic, Hydrocolloid Process, \\ Human Mometasone, Mometasone, Container, Olopatadine, Viscosity, \\ Medium Chain, Alcohol, Dose, Nasal, Linaclotide, Lipid, Oil Tricaprylate, Affected, \\ Aqueous, Topical, Modifying, Application Shear, Wax Tetracycline.\end{tabular}                                                                                & 43                         \\ \hline
13             & \begin{tabular}[c]{@{}l@{}}New formulations based in coated \\ or granular vehicles\end{tabular}                                   & \begin{tabular}[c]{@{}l@{}}Active, Ingredient, Layer, Agent, Coated Bead, Pharmaceutical, Distal Intestine, \\ Bead Granule, Granule Layer, System Stimulant, Inter Coated, Tablet, Disorder, \\ Central Nervous, Sufficient, Solvent, Portion, Oral, Non-Volatile, Vehicle, \\ Attention Deficit, Release, Immediate, Water, Administration, Embodiments, \\ Cancer, Core, Components, Abuse.\end{tabular}                                                                                                             & 195                        \\
\bottomrule  
\end{tabular}
\end{table*}

\begin{table*}[htb]
\centering
\scriptsize
\caption{Overview of the Topics (14-24) built by PATopics framework, its words, title, and the number of patents.}
\begin{tabular}{llll}
\toprule
\textbf{Topic} & \textbf{Topic Title}     & \textbf{Words}           & \textbf{Number of patents} \\ \midrule
14             & Chronic pathologies                                                                                                                & \begin{tabular}[c]{@{}l@{}}Colon Continuity, Bowel Syndrome, Patients, Glp, Agonist Teduglutide, \\ Intestinal Absorption, Receptor, Treatment, Cocaine Hydrochloride, Uses, Novel, \\ Diabetes, Peptides, Therapy, Caprylic, Hydroxybenzoyl Amino, Nausea Vomiting, \\ Regimes, Medicine, Hepatic Impairment, Middle Ear, Primary Package, \\ Package Insert, Chemotherapy Radiotherapy, Etc, Events, Ethylamino, \\ Euglobulin, Eva, Eudragit.\end{tabular}                                                           & 34                         \\ \hline
15             & \begin{tabular}[c]{@{}l@{}}Elethronic devices and medicine \\ administration devices\end{tabular}                                  & \begin{tabular}[c]{@{}l@{}}Medicament, Member, Housing, Device, Container, Needle, Position, Injection, \\ Apparatus, Energy Storage, Chamber, Force, Actuator, Fluid, Syringe, Naloxone, \\ Longitudinal Axis, Dose, Gas, Systems, Antagonists, Opening, Surface, Movable, \\ Distal, Electronic Circuit, Reservoir, Portion Puncturer, Relative, Jet Injector.\end{tabular}                                                                                                                                           & 235                        \\ \hline
16             & \begin{tabular}[c]{@{}l@{}}Methods of administration, \\ abuse-deterrent, controlled release \\ and stability related\end{tabular} & \begin{tabular}[c]{@{}l@{}}Drug, Water, Soluble, Products, Abuse Deterrent, Administration, Transmucosal, \\ Likelihood Improper, Physical Integrity, Example Blade, Delivery, Colchicine, \\ Ion Exchange, Liposome, Resin, Formulation, Release, Embodiments, \\ Microparticles, Coating, Nasal, Polymer, Small Volume, Opioids Drugs, \\ Organic, Adverse Physiological, Preferred Embodiment, Combinations Dpp, \\ Ways, Symptoms Inventive.\end{tabular}                                                           & 197                        \\ \hline
17             & \begin{tabular}[c]{@{}l@{}}Formulations for gastrointestinal, \\ hepatic and dermatological uses\end{tabular}                      & \begin{tabular}[c]{@{}l@{}}Acid, Amino, Methyl, Tranexamic, Therewith, Salt, Carboxylic, Solutions, \\ Phenyl Imidazol, Dsrna, Dermatological, Fenofibric, Mixture, Related, \\ Obeticholic, Cyclopropanecarboxamido Benzoic, Difluorobenzo Dioxol, \\ Moieties, Compositions, Crystals, Gene, Naphthoic, Synthesis Deoxycholic, \\ Origin Pyrogenic, Citric, Cyclic Boronic, Bile Acids, Oxo Pyrazol, \\ Bis Monoethanolamine, Aberrant Uric.\end{tabular}                                                             & 240                        \\ \hline
18             & Genetic Therapy and related                                                                                                        & \begin{tabular}[c]{@{}l@{}}Exon Skipping, Capable Target, Antisense Oligonucleotides, Site Exon, \\ Skipping Dystrophin, Gene Forth, Seq, Methods, Dystrophin Gene, \\ Muscular Dystrophy, Pathologies Antisense, Molecule Capable, Forth Seq, \\ Target Site, Human, Eudragit, Euglobulin, Event, Estrogenic, Ethylvinyl, \\ Evaluation, Everolimus, Example, Estradiol Valerate, Ethanesulfonate, \\ Ethenocycloprop, Ethoxy Methoxyphenyl, Etelcalcetide, \\ Etc1002, Zwitterionic.\end{tabular}                     & 23                         \\ \hline
19             & \begin{tabular}[c]{@{}l@{}}Analgesics and related receptors \\ modulators\end{tabular}                                             & \begin{tabular}[c]{@{}l@{}}Opioid, Receptor, Antagonist, Modulators, Products, Naloxone, \\ Abuse Resistant, Pain, Nasal, Matrix, Level Euphoric, Accumulation,\\ Nature Table, Ligands, Symptoms Inventive, Peripheral, Polymer, Effective, \\ Amounts, Regimen Sedation, Analgesics, Disorders, Pre, Tablets, \\ Time, Dose, Peptide Amides, Immediate, Formula, \\ Inhibition Penetration.\end{tabular}                                                                                                              & 104                        \\ \hline
20             & \begin{tabular}[c]{@{}l@{}}Gene therapy and \\ new formulations applied in cancer \\ and viral diseases\end{tabular}               & \begin{tabular}[c]{@{}l@{}}Therapeutic, Agent, Presence Acidic, Ingenol Angelate, Potent Anticancer, \\ Aprotic Solvent, Buffer, Compositions, Methods, Example, Liposome, \\ Administration, Sirna, Antiviral, Hcv, Membrane, Hyperactivity Disorder,  \\ Attention Deficit, Hours Longer, Additional, Bioavailability, Cyclic Boronic, \\ Mammal, Polyethylene Glycol, Application, Hydrophobic, Oil, Active, \\ Surfactant Vegetable, Prostaglandin Travoprost.\end{tabular}                                         & 139                        \\ \hline
21             & \begin{tabular}[c]{@{}l@{}}Dispersions and \\ new pharmaceutical formulations\end{tabular}                                         & \begin{tabular}[c]{@{}l@{}}Solid, Forms, State, Carboxamide, Ethyl Imidazo, Pyrrolo Pyrazin, \\ Trifluoroethyl Pyrrolidine, Dispersion, Synthesis Products, Processes, \\ Rheumatoid Arthritis, Bis Hydroxyphenyl, Dihydro Oxoquinoline, \\ Tacrolimus Hydrophilic, Hydrochloride, Vehicle, Water Miscible, \\ Bioavailability, Bendamustine, Dosage, Tigecycline, Kits, \\ Ambient Temperature, Therewith, Cyclopropanecarboxamido Benzoic, \\ Pharmaceutical, Itraconazole, Antiviral, Oral, Vitamin.\end{tabular}    & 165                        \\ \hline
22             & Brain disorders                                                                                                                    & \begin{tabular}[c]{@{}l@{}}Amphetamine, Abuse Liability, Bioavailability Amphetamine, \\ Hyperactivity Disorder, Attention Deficit, Parenteral Routes, \\ Intravenous Intranasal, Adhd Narcolepsy, Doses, Alternative, Obesity, \\ Oral, Chemical Moiety, Compounds, Prodrugs, Administration, Disorders, \\ Method, Treatment, Lysine, Hours Longer, Night, Ascending, Potential, \\ Coated, Masked, Rate, Therapeutic, Salt, Population Beads.\end{tabular}                                                           & 24                         \\ \hline
23             & \begin{tabular}[c]{@{}l@{}}Ophthalmic conditions and \\ eye/mucus nanoparticulate \\ delivery systems\end{tabular}               & \begin{tabular}[c]{@{}l@{}}Particles, Agents, Applications, Ophthalmic Applications, Embodiments, \\ Eye, Body Wide, Spectrum Applications, Mucus Barriers, Surface Coatings, \\ Imaging Diagnostic, Efficient Transport, Compositions, Solubility, \\ Instances Ophthalmic, Methods, Delivery, Pharmaceutical, Aqueous, \\ Suspension Medium, Drug, Megestrol, Nanoparticulate Megestrol, \\ Pulmonary, Polymeric Carriers, Minimal Polymeric, Routes, Lipid, \\ Snalp, Opioid Dependency.\end{tabular}                & 85                         \\ \hline
24             & Methods of monitoring and image                                                                                                    & \begin{tabular}[c]{@{}l@{}}Nitrogen Scavenging, Ammonia Blood, Monitoring Nitrogen, \\ Level Measurement, Technique, Scavenging Drugs, Disclosure, Drug, \\ Methods, Therapeutic, Dosage, Inter Alia, Ethyl Eicosapentaenoate, \\ Mixed Dyslipidemia, Itraconazole, Oxymetazoline, Neurological, \\ Metastasis Recurrence, Facbc, 18f Facbc, Imaging, \\ Extrapyramidal Reactions, Ethylpyridie, Euglobulin, Eudragit, \\ Ethoxymethyl Phosphonic, Eva, Event, Everolimus, Example Blade.\end{tabular}                  & 24                         \\
\bottomrule
\end{tabular}
\end{table*}

\begin{table*}[htb]
\centering
\scriptsize
\caption{Overview of the Topics (25-29) built by PATopics framework, its words, title, and the number of patents.}
\begin{tabular}{llll}
\toprule
\textbf{Topic} & \textbf{Topic Title}     & \textbf{Words}           & \textbf{Number of patents} \\ \midrule
25             & \begin{tabular}[c]{@{}l@{}}Crystalline and \\ amorphous formulations\end{tabular}                                                  & \begin{tabular}[c]{@{}l@{}}Crystalline, Forms, Polymorphic, Salts, Rifaximin, Inhibitor, Amorphous, \\ Novel, Methyl, Processes, Phenyl, Potent, Hydrogen Sulfate, \\ Adverse Physiological, Preparations, Ways, Bis Trifluoromethyl, Events, \\ Pyrazolo Pyrimidin, Production Medicinal, Alcohol Crystallization, \\ Hot Raw, Content Object, Time Drying, Treatment, Diseases, Addition, \\ Water, Pharmaceutical, Determinate Temperature.\end{tabular}                                                             & 197                        \\ \hline
26             & Hormonal related                                                                                                                   & \begin{tabular}[c]{@{}l@{}}Progesterone, Micronized Progesterone, Estradiol Micronized, \\ Natural Hormone, Replacement, Therapies Estrogen, Formulations, \\ Ultra, Example, Administering, C12, Medium Chain, Comprises, \\ Oil, Pharmaceutical, Embodiments, Need, Monolithic Intravaginal, \\ Ethylester, Eudragit, Euglobulin, Eva, Event, Everolimus, \\ Zwitterionic, Ethoxymethyl Phosphonic, Esters, Etc, \\ Etelcalcetide, Ethanediamide.\end{tabular}                                                        & 27                         \\ \hline
27             & Ophthalmic and nasal formulations                                                                                                  & \begin{tabular}[c]{@{}l@{}}Aqueous, Solution, Ophthalmic, Weight, Bimatoprost Ophthalmic, \\ Ppm, Rpm, Benzalkonium Chloride, Hypertension Thereto, \\ Glaucoma Ocular, Liquid, Method, Administration, Water, \\ Suspension, Salt, Sufficient, Non, Organic, Cyanocobalamin, Agent, \\ Carrier, Cyclodextrin, Olopatadine, Rhinitis, Bioavailability, \\ Brimonidine Topical, Intranasal, Spironolactone, Pressure Eyes, \\ Elevated Intraocular.\end{tabular}                                                         & 206                        \\ \hline
28             & Modulators related                                                                                                                 & \begin{tabular}[c]{@{}l@{}}Modulators Atp, Cystic Fibrosis, Transmembrane Conductance, \\ Abc Transporter, Cassette Abc, Regulator Cftr, Diseases, Compounds, \\ Methods, Cftr, Indole, Ethylester, Eudragit, Evaporation, Euglobulin, \\ Evasion, Event, Everolimus, Example, Estradiol Valerate, Estrogenic, \\ Etc, Etelcalcetide, Ethanediamide, Ethoxy Methoxyphenyl, Zwitterionic, \\ Expenditure, Expression Alas1, Extended Maximal, Exterior.\end{tabular}                                                     & 23                         \\ \hline
29             & Cardiovascular related                                                                                                             & \begin{tabular}[c]{@{}l@{}}Risk Cardiovascular, Therapy, Eicosapentaenoic, Event, Derivative, \\ Treating Cardiovascular, Methods, Hypertriglyceridemia Embodiments, \\ Ethyl, Ester, Acid, Lipid, Embodiments, Disease, Level, Related, Blood, \\ Need, Pharmaceutical, Triglycerides, Pirfenidone, Kinase Activators, Cancer, \\ Pyruvate Kinase, Immunomodulatory, Concomitant, Ldl, Injectable, \\ Suitable, Anti Emetic.\end{tabular}                                                                              & 79  \\  
\bottomrule
\end{tabular}
\end{table*}

Some topics are highly generic, such as Topic 0 (249 patents), Topic 1 (421 patents), Topic 11 (248 patents), and Topic 21 (165 patents), encompassing a range of patents that correlate at some point but belong to the distinct pharmaceutical areas. For example, in Topic 11, entitled - General formulations, methods, and new administration forms and devices, a tablet formulation containing a hypoglycemic agent are grouped with a patent about a capsule of an analgesic. This happens because both are related to – formulations (pharmaceutical forms) but are from different drug classes. Otherwise, some topics are much more specific, such as Topic 10, entitled - Dermatological conditions and treatments, in which all 175 patents are topical/transdermal pharmaceutical formulations for skin diseases; Topic 19, entitled - Analgesics and related receptors modulators, in which 104 patents are related to analgesics and their interaction with pain receptors. It is interesting to note that PATopics always finds a way to group pharmaceutical patents in a proper context. However, some contexts are more generic than others. Usually, topics with a significant number of patents consequently group them using a generic intersection between them.\looseness=-1

Also, we observe in PATopics the correlation between the patents and the companies who patented and the molecules involved in the patents. We indicated in Figure~\ref{fig:3}-B and Figure~\ref{fig:3}-C the top 10 companies holding pharmaceutical patents and molecules, respectively.\looseness=-1

Allergan and Novartis are the two companies that patent the most ($\sim$130 patents each), followed by Takeda. These three companies hold almost twice as many patents as the other top 10 companies. Each one has its specific interests and based on patents, PATopics is able to identify, by topic, the most engaged companies.\looseness=-1

In Figure~\ref{fig:3}-D, we chose, as examples, four Topics to discuss: The Topic 1 and 8 do not present the dominance of any specific company, having their patents distributed among many companies. Topic 10 shows dominance by Galderma, who holds 37 patents on this Topic, followed by 18 patents of Horizon. Topic 29 illustrates a Topic in which there is a majority portion of patents owned by a single company – 42 out 79 patents belonging to Amarin Pharma.\looseness=-1

They sum together 410 patents which the Icosapent Ethyl is the most patented molecule, responsible for 67 patents. Table 3 summarizes these ten molecules, their pharmaceutical details, and the main claim about them. They are three patent-free molecules, epinephrine, testosterone, and nitric oxide, and seven are patented by companies, of which two are still in patent time and the other five the patent expiration.\looseness=-1

\begin{table*}[htb]
\centering
\caption{Top 10 patented Molecules, their Topics, pharmaceutical information, and main claim.}
\resizebox{1.0\textwidth}{!}{
\begin{tabular}{llllllll}
\toprule
\textbf{Molecule}                                                        & \textbf{\begin{tabular}[c]{@{}l@{}}First Brand \\ name\end{tabular}} & \textbf{\begin{tabular}[c]{@{}l@{}}Patent \\ Company \end{tabular}}                                             & \textbf{\begin{tabular}[c]{@{}l@{}}Pharmaceutical \\ Class\end{tabular}}                                                    & \textbf{\begin{tabular}[c]{@{}l@{}}Year of Patent / \\ discovered\end{tabular}} & \textbf{Indication}                                                                                       & \textbf{\begin{tabular}[c]{@{}l@{}}Topic \\ (Number of \\ patents per Topic)\end{tabular}}                                                & \textbf{Main patent claims}     \\ \midrule
\begin{tabular}[c]{@{}l@{}}Icosapent \\ Ethyl\end{tabular}               & Vascepa\textregistered                  & \begin{tabular}[c]{@{}l@{}}Amarin \\ Pharmaceuticals\end{tabular}   & Antihyperlipidemic                                                               & 2012                                 & \begin{tabular}[c]{@{}l@{}}Severe \\ Hypertriglyceridemia\end{tabular}                                    & \begin{tabular}[c]{@{}l@{}}T5 (1), T12 (18), \\ T14 (1), T24 (5), \\ T29 (42)\end{tabular}                                                & \begin{tabular}[c]{@{}l@{}}Stable compositions, Methods of treating \\ Hyperlipidemia, Methods of reducing \\ the risk of a cardiovascular event.\end{tabular}   \\ \hline
Aripiprazole                                                             & Aristada\textregistered                 & Alkermes plc.                                                       & Antipsychotic                                                                    & 2015                                 & Schizophrenia                                                                                             & \begin{tabular}[c]{@{}l@{}}T4 (6), T5 (6), \\ T8 (12), T9 (2), \\ T13 (7), T15 (5), \\ T16 (6), T17 (1), \\ T20 (1), T21 (1)\end{tabular} & \begin{tabular}[c]{@{}l@{}}Controlled activation, low hygroscopic \\ preparation, in-body device, multi-mode \\ communication and highly reliable \\ ingestible event markers and methods \\ for using the same, Implantable \\ zero-wire communications \\ system, derivates, methods of \\ administration.\end{tabular} \\ \hline
Epinephrine                                                              & Patent-free               &                                                                     & \begin{tabular}[c]{@{}l@{}}Alpha- and \\ beta-adrenergic \\ agonist\end{tabular} & \begin{tabular}[c]{@{}l@{}}Discovered \\ in 1896 \end{tabular}                   & \begin{tabular}[c]{@{}l@{}}Emergency \\ treatment of \\ type I allergic \\ reactions.\end{tabular}        & \begin{tabular}[c]{@{}l@{}}T5 (1), T8 (6), \\ T11 (3), T15 (32), \\ T16 (1), T20 (3)\end{tabular}                                      & \begin{tabular}[c]{@{}l@{}}Formulations, stable suspensions, automatic \\ injector, devices, eletronic devices, more \\ potent and less toxic formulations, \\ Apparatus for delivery.\end{tabular}                                                                                                                       \\ \hline
\begin{tabular}[c]{@{}l@{}}Methylphenidate \\ hydrochloride\end{tabular} & Ritalina\textregistered                 & \begin{tabular}[c]{@{}l@{}}Novartis / \\ Celgene \end{tabular}        & \begin{tabular}[c]{@{}l@{}}Central Nervous \\ System stimulant\end{tabular}      & 2002                                 & \begin{tabular}[c]{@{}l@{}}Attention Deficit \\ Hyperactivity \\ Disorders \\ (ADHD)\end{tabular}         & \begin{tabular}[c]{@{}l@{}}T4 (29), T13 (12), \\ T27 (1) \end{tabular}                                                                                                             & \begin{tabular}[c]{@{}l@{}}Methods of administration and production, \\ compositions, extended release, chewable \\ tablet, suspensions, powder and aqueous \\ suspension.\end{tabular}                                                                                                                                   \\ \hline
\begin{tabular}[c]{@{}l@{}}Hydrocordone \\ bitartrate\end{tabular}       & Vicodin\textregistered                  & Abbvie                                                              & \begin{tabular}[c]{@{}l@{}}Narcotic \\ analgesic\end{tabular}                    & \begin{tabular}[c]{@{}l@{}}Discovered \\ in 1923 \end{tabular}                   & \begin{tabular}[c]{@{}l@{}}Severe and \\ chronic pain\end{tabular}                                        & \begin{tabular}[c]{@{}l@{}}T2 (20), T4 (14), \\ T13 (8), T27 (2)\end{tabular}                                                          & \begin{tabular}[c]{@{}l@{}}Immediate release, treating pain in patients \\ with hepatic impairment, abuse resistant \\ drug formulation, Abuse-proofed dosage \\ form, Tamper resistant dosage forms.\end{tabular}                     \\ \hline
Ibrutinib                                                                & Imbruvica\textregistered                & \begin{tabular}[c]{@{}l@{}}Janssen and \\ Pharmacycles\end{tabular} & \begin{tabular}[c]{@{}l@{}}Kinase \\ inhibitor\end{tabular}                      & 2013                                 & Lymphoma (MCL)                                                                                            & T5 (3), T9 (37)                                                                                                                        & \begin{tabular}[c]{@{}l@{}}Clinical methods of treating and preventing \\ graft versus host disease, uses, Crystalline \\ forms, formulations.\end{tabular}   \\ \hline
Nitric Oxide                                                             & Patent-free               &                                                                     & Vasodilator                                                                      & \begin{tabular}[c]{@{}l@{}}Discovered \\ in 1772 \end{tabular}                   & \begin{tabular}[c]{@{}l@{}}Congenital cardiac \\ disease and cases of \\ respiratory failure\end{tabular} & \begin{tabular}[c]{@{}l@{}}T5 (1), T8 (27), \\ T20 (1), T21 (4), \\ T29 (2)\end{tabular}                                                  & \begin{tabular}[c]{@{}l@{}}Methods for inhaled treatment, Methods \\ of reducing the risk of occurrence of \\ pulmonary edema in children, gas delivery \\ device, devices, kit for conversion NO2 in \\ NO, apparatus for monitoring, storage \\ cassette.\end{tabular}      \\ \hline
Testosterone                                                             & Patent-free               &                                                                     & Hormone                                                                          & \begin{tabular}[c]{@{}l@{}}Named \\ in 1935 \end{tabular}                        & Hormone reposition                                                                                        & \begin{tabular}[c]{@{}l@{}}T3 (3), T4 (2), \\ T5 (10), T8 (5), \\ T10 (9), T13 (4)\end{tabular}                                           & \begin{tabular}[c]{@{}l@{}}Gel composition, transdermal delivery, \\ controlled release, nasal application, \\ spreading implement, compositions.\end{tabular}  \\ \hline
\begin{tabular}[c]{@{}l@{}}Sumatriptan \\ Succinate\end{tabular}         & Imitrex\textregistered                  & \begin{tabular}[c]{@{}l@{}}GlaxoSmithKline \\ (GSK)\end{tabular}    & \begin{tabular}[c]{@{}l@{}}Selective serotonin \\ receptor agonists\end{tabular} & 1982                                 & Migraines                                                                                                 & \begin{tabular}[c]{@{}l@{}}T2 (1), T3 (1), \\ T5 (1), T8 (16), \\ T11 (1), T15 (7), \\ T16 (4)\end{tabular}                               & \begin{tabular}[c]{@{}l@{}}Multilayer dosage forms, formulation, \\ Casing, Needleless injector, transdermal \\ method, iontophoretic administration, \\ nasal delivery, nasal devices.\end{tabular}        \\ \hline
\begin{tabular}[c]{@{}l@{}}Migalastat \\ Hydrochloride\end{tabular}      & Galafold\textregistered                 & \begin{tabular}[c]{@{}l@{}}Amicus \\ therapeutics\end{tabular}      & \begin{tabular}[c]{@{}l@{}}Miscellaneous \\ metabolic \\ agents\end{tabular}     & 2021                                 & Fabry disease                                                                                             & T7 (25), T9 (4)                                                                                                                        & \begin{tabular}[c]{@{}l@{}}Methods of treatment and prediction, \\ Dosing regimens for the treatment \\ of lysosomal storage diseases.\end{tabular}         \\
\\ \bottomrule
\end{tabular}
}
\end{table*}

We identified the main subjects of these 4,832 pharmaceutical patents and also the main molecules and companies. The main subjects (Figure~\ref{fig:3}-E) contained in the patents were in descending order of representativeness: formulations and compositions, new compounds and prodrugs, chronic conditions, pain, clinical methods, devices, viral and cancer-related, dermatological, gastrointestinal, gene therapy, brain disorders, ophthalmic and nasal. As expected, we observed many patents regarding new formulations, compositions, new compounds, and synthesis of prodrugs since most patents have formulation and composition terms in their descriptions. The Topic Modeling strategy tries to gather the patents according to their descriptions. Since the patent description is free-text, we observed that some patents have more detailed information than others. This behavior may cause the Topic Modeling strategy to build a more generic topic.\looseness=-1

There were not many patents as expected regarding viruses and cancer, but perhaps future data after the Covid-19 pandemic, these data will change, given the investment in antivirals. However, it is important to mention the long research periods required to obtain new molecules and/or effective formulations. Furthermore, the long patenting time often leads researchers to disregard patenting, preferring to publish their results in peer-reviewed journals and conferences.\looseness=-1

The third most patented subject covers chronic diseases (Figure~\ref{fig:3}-F) such as hypertension, cardiovascular-related, and diabetes. Indeed, this subject concentrates on the substantial profits of the pharmaceutical industries~\citep{Reinhardt2001, Waters2018}.\looseness=-1

One hundred million Americans have dyslipidemia, a cholesterol imbalance, indicating high levels of low-density lipoprotein (LDL) cholesterol or low levels of high-density lipoprotein (HDL) cholesterol. 80 million individuals live with hypertension, which is a significant risk factor for heart disease and stroke. 10.9\% of the adult population lived with type 2 diabetes. Hypertension, dyslipidemia, and diabetes type 2 in sum represented about 1.7 million dollars of costs per year in the U.S.~\citep{Waters2018}. Therefore, increasing patents involving products for chronic diseases are expected, as well as new molecules applied to this subject.\looseness=-1

\subsection{Contributions}
This section was structured to highlight the framework's main contributions in a general context, showing contributions to potential users. Thus, from the general contributions, the results were presented in specific inputs from three scenarios (a) Researchers who work with patents, (b) Chemists who develop patents and (c) Companies or industries who want to use or buy patents.\looseness=-1

\subsubsection{General contributions}
Concerning pharmaceutical patents, one of the main drawbacks is the decentralization of patent data around the world. Each country has its own patent repository, or even institutions within countries deposit these patents by themselves in their own repository. In this way, when there is a search to consult patents, it is challenging to find patents that faithfully represent the query. There are no large patent repositories that are structured to cover a large number of patents on different topics. For this reason, the main general contribution of the PATopics framework is its ability to centralize patents, given a search query used to fetch the patents, in a single environment of search.\looseness=-1

In addition, PATopics not only gathers patents but also manages to group them into topics based on their similarities, which makes searching by topic even easier. This summarization of topics from patents in the pharmaceutical field manages to point out the main patented subjects, highlights the most relevant topics, directs the interest of companies regarding the scientific theme, highlights more patented molecules as well as notifies the evolution of pharmaceutical forms, packaging, devices, methods, among others.\looseness=-1

The idea of bringing together patents from a large area, such as pharmaceuticals in a framework facilitates access to data and makes it possible to discuss and compare patents developed in different places by different companies. The discussion and comparison, in this case, allow a great evolution of the creation and innovation process, where, through the tool, it is possible to observe the evolution over time, numerically expressed and easily interpreted. In this way, PATopics represents not only a tool where there can be search but also where there is data analysis, similarities, evolution in the timeline, growth of patenting by companies, and identification of areas and possibilities.\looseness=-1

\subsubsection{Specific contributions}
\begin{figure*}[htb]
	\centering
	\includegraphics[scale=0.35]{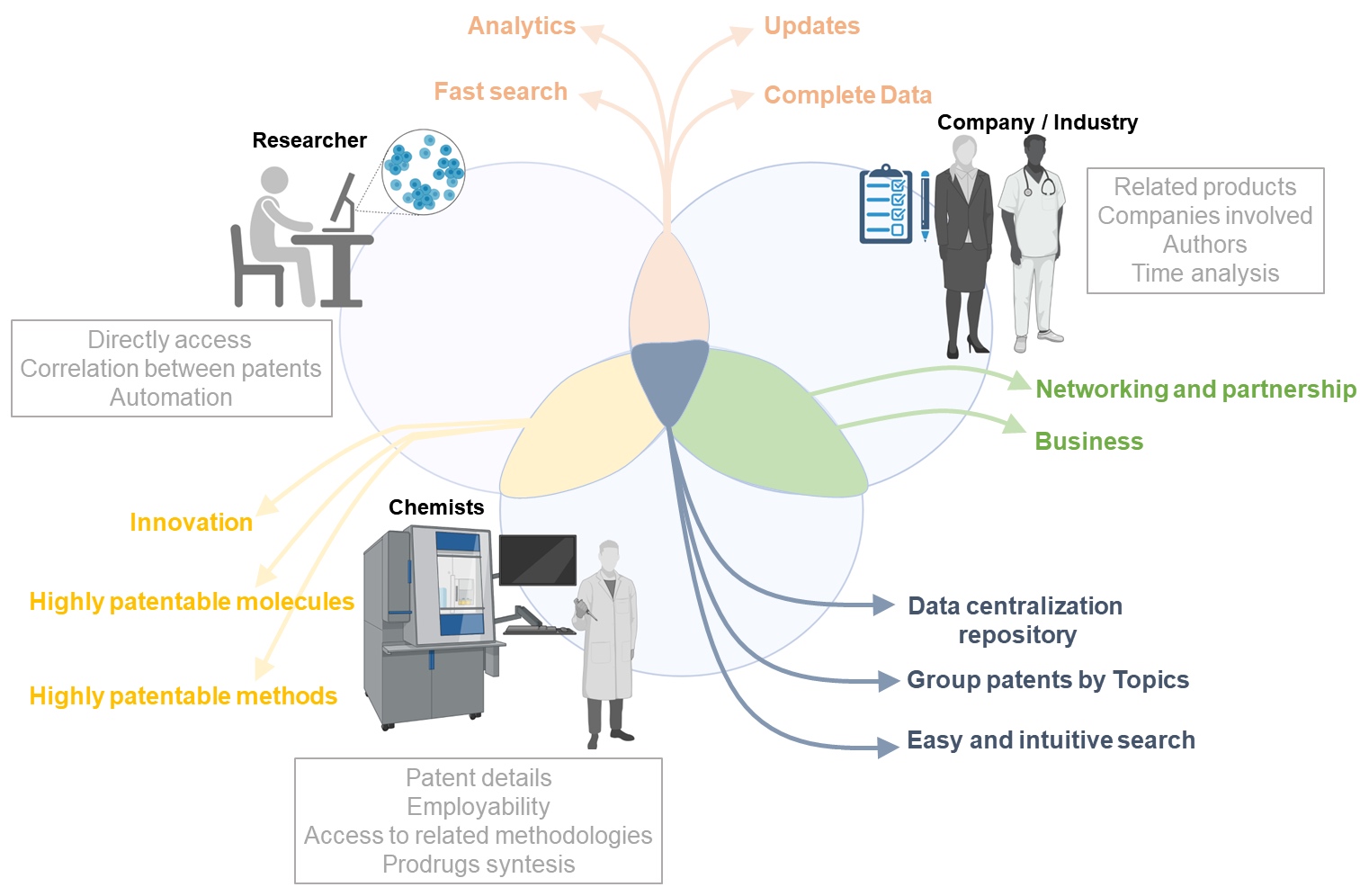}
	\caption{The profile of potential users of PATopics framework and their main interest that the framework can engage in. The first profile is researchers who work with patents and their studies. The second is the chemists, who develop the patents and the third is companies and industries, who buy or use patents.}
	\label{fig:4}
\end{figure*}

This section presents three user scenarios (Figure~\ref{fig:4}) to highlight the contributions of the PATopics framework to different user profiles with different goals. The first scenario is the users who research patents, represented by academic researchers or researchers from pharmaceutical industries, both interested in seeking patents and performing analyzes based on their claims.\looseness=-1

From the point of view of a researcher user, the PATopics framework is an environment capable of providing direct access to patents, being possible to be redirected via a link to the domain in which they were deposited. The tool provides the functionality to correlate patents with each other. This functionality will be covered in case study 04 in the next Section. In addition, the possibility of working in a fully automatic environment, where patents have already been grouped into topics that correlate them, facilitates the researcher's work, optimizing working time.\looseness=-1

The common interests between the researcher user and the company and industry user are the possibility of carrying out fast searches, in a standardized environment, following updates of existing patents, accessing complete patent data ranging from their original text to data on inventors, company responsible for patents, active molecules involved, pharmaceutical form or device, administration method, dosages, and drugs concentration. Furthermore, have access to analytics that will contribute to decision-making regarding particular research, a possible purchase, or the use of a particular patent.\looseness=-1

A company or industry interested in patents, in the use or purchase of their domains, may benefit from using a platform such as PATopics, since access to related products allows a direct comparison between products, as well as obtaining information from companies involved in patentability. Also, access to patent information from companies in the same niche helps develop portfolios of competing products in the market, as well as innovation. Access to inventors' data can directly contribute to hiring employees focused on a particular development interest. The tool can also help in timeline studies, where it can follow patent trends, helping to implement purchase processes for domains of innovative products in the market.\looseness=-1

It is important to mention that between companies/ industrial users and patent developers’ users, there is a direct relationship of mutual gain based on the construction of networks, which will enable a partnership. A company will not always be interested in purchasing patents. In many cases, there is a business dynamic between both sides, who produces and uses them to generate shared profits.\looseness=-1

Finally, for this third user, patent developers, usually chemists, access to detailed patent data in a simple way helps develop products in the same niche and similar methodologies. Nowadays, many patents are dedicated to synthesizing prodrugs, and access to these patents is extremely important to the chemist responsible for syntheses. The knowledge of chemical intermediates used, as well as synthesis conditions and the use of catalysts, help to develop increasingly optimized methodologies with high yields and purity.\looseness=-1

It is important to mention that for these developers, access to companies involved in patentability helps in the employability process, where the company is identified by area of interest and expertise.\looseness=-1

The common point between researchers and developers is regarding access to innovation, in addition to the easy identification of highly patentable molecules and methods. Pharmaceutical patents evolve over the years and in this way, a formulation containing a specific molecule, application device, or pharmaceutical form does not continue to be patented if it is not highly profitable. Disruptive innovations are increasingly present in the pharmaceutical niche, and access to this data in a standard framework, to our best knowledge, was done for the first time in PATopics.\looseness=-1

\section{PATopics validation}
In this section, we will address some case studies, which are nothing more than examples of actual applications of the PATopics framework, showing their results in a real scenario, mimicking its use by one of the users mentioned in section -- Specific contributions.\looseness=-1

\subsection{Case study 01 – Topic study}
The study of a Topic can be carried out by a researcher user interested in studying patents on a particular topic. This way, the user will initially select the topic of interest. For instance, we selected the Topic 8.\looseness=-1

Topic 8 was composed of 30 words: Delivery, System, Nitric Oxide, Gas, Transdermal, Surface, Nasal, Vehicle, Active, Nosepiece, Viscosity, Pulmonary, Unit, Administration, Polyorthoester Polymer, Buprenorphine Sustained, Polar Aprotic, Agent, Sustained, Conversion Nitrogen, Dioxide NO2, Antioxidant Ascorbic, Exemplary Gas, Solvent, Inhalation, Estrogen, Data, Method, Flowable, Valve Assembly. Based on these works, we entitled this Topic as – Delivery systems and devices.\looseness=-1

Topic 8 was composed of 195 patents and the main themes of these patents were: related to the sustained and controlled release of drugs; pulmonary, nasal, and transdermal delivery systems; long-acting delivery system; Iontophoresis; aerosols; delivery of ocular implants; osmotic drug delivery systems; implants; nanomedicines; packaging system; and devices such as piston-pumping system, system for birth control, powder delivery devices, dose counter for inhalers, body-associated receiver, user-activated devices, abuse-resistant devices, mucoadhesive devices, ingestible event marker system.

This topic encompasses the most significant innovations in the pharmaceutical field, where innovation occurs either in terms of formulations capable of overcoming the most significant drawbacks of conventional pharmaceutical forms or in developing devices that optimize the use and application of the pharmaceutical product. Delivery systems are formulations that improve the application of drugs, genes, proteins, and peptides~\citep{VIEGAS2022103964}. These delivery systems are able to control the release of active agents, modulate the action time, be successful in the targets, are biocompatible, and are less toxic~\citep{Gall2022, Iqbal2018, Reinholz2018, Rosa2018}. The use of devices increasingly expands a pharmaceutical formulation's capacity, overcoming the pharmacological capabilities of this same passively applied formulation. Pulmonary and nasal applications, for example, are highly enhanced with the use of devices\citep{Abdelaziz2018, Deepika2019}. Analgesics and medications for continuous use have their pharmacodynamics controlled using automated application devices. In brief, this topic deals with patents that have developed enhancement systems, where the end result is always improvement in the patient's therapy and quality of life~\citep{Liu2016}.\looseness=-1

Therefore, the user can verify at the end of the Topic 8 analysis that the universe of delivery systems is something broad and versatile, but some words clearly indicate the prevalence with emphasis on physical methods such as iontophoresis, formulations aimed at ocular use, and nanomedicine.\looseness=-1

\subsection{Case study 02 – Company study}
In the study on companies, we can relate to a user Company ``Y'' interested in discovering what has been patented by its competing company in the market. For this, we selected the company GlaxoSmithKline (GSK). GSK has 32 collected patents involving 67 inventors. This company is very versatile, having patents in several topics. Its oldest patent collected is from 2008 -- Phenethanolamine derivatives for treatment of respiratory diseases, belonging to Topic 11 (General formulations, methods, and new administration forms and devices), and the newest is from 2021 -- Niraparib compositions, belonging to Topic 21 (Dispersions and new pharmaceutical formulations).\looseness=-1

GSK also has patents regarding nicotine lozenge compositions, non-steroidal anti-inflammatory drugs, decongestants, and anti-histamines compositions from Topic 2 (Pain treatments and related to chronic diseases), formulations of salts from Topic 1 (General molecules-based inhibitors, prodrugs, and modulators), inhalation devices delivering phenylethanolamine derivatives from Topic 6 (New compounds and prodrugs), and patents from Topic 9 related to cancer, Topic 11 related to respiratory diseases, Topic 13 about sleep disturbances, Topic 27 with a patent of the composition of suspensions.
The dominant Topics of GSK are Topic 15 (Electronic devices and medicine administration devices), with ten patents and Topic 19 (Analgesics and related receptors modulators), with six patents. In Topic 15, the main subject adopted by GSK is devices acting as medicament dispensers and the period of these patents is from 2009 to 2016. In Topic 19, GSK patented muscarinic receptor antagonists and recent patents aiming for combinations of muscarinic receptor antagonists and a beta-2 adrenoreceptor agonist.\looseness=-1

Therefore, Company ``Y'' may conclude, based on PATopics' analysis of GSK, that they were a company with several interests, acting in various pharmacological and technological scenarios. GSK has a constant number of patents over the years, indicating a constant concern with innovation.\looseness=-1

Finally, the study of the Company using PATopics' results allowed us to observe the GSK as a consistent company with a broad portfolio of products and to study the main interests of the company. PATopics results demonstrate the ability of this framework to perform a complex, multifactorial interpretation of companies with a wide range of products dedicated to different pharmaceutical subareas.\looseness=-1

\subsection{Case study 03 – Molecule study}
The molecule case of study may be performed by a chemist user who aims to study patents related, for example, to antidiabetic molecules to explore new pharmaceutical technologies with them. For this, the molecule adopted in this case study is Metformin hydrochloride, which has been adopted as therapy for diabetes mellitus.\looseness=-1

The metformin's mechanism of action is based on hepatic gluconeogenesis inhibition and opposing the action of glucagon, redirecting the cell's energy metabolism. However, in the last years, Metformin has gained attention in tumorigenesis, by indirect effects of systemic reduction of insulin levels, and directly, by induction of energetic stress; but these effects require further investigation\citep{Eibl2021, Pernicova2014}.\looseness=-1

The PATopics identified metformin at 29 patents, in which the newest is a patent of Boehringer Ingelheim -- Dpp-Iv Inhibitor combined with a further antidiabetic agent, tablets comprising such formulations, their use, and process for their preparation. This patent was filed in 2019 and granted in 2021 and PATopics classified as Topic 16 (Methods of administration, abuse-deterrent, controlled release and stability related) as the dominant topic but also determined a correlation with Topic 11 (General formulations, methods, and new administration forms and devices) and Topic 9 (Inhibitors related to viral and cancerous process).\looseness=-1

Metformin has patents in 10 Topics being Topic 16 the majority, with 12 patents. All the patents on this topic are about combinations with other drugs, formulations, and uses. The drugs used in combination with metformin in this topic are Empagliflozin, Linagliptin, and Pioglitazone Hydrochloride, and these patents belong to Boehringer Ingelheim and Takeda.\looseness=-1

An interesting patent of metformin is -- Extended-Release Suspension Compositions from Sun Pharm company, in which the inventors claim a suspension containing metformin that sustains its release for at least seven days. The formulation comprises multiple coated cores. This patent was filed in 2016 and granted in 2018 and the PATopics classified as Topic 4 (Formulations with the controlled, sustained, or extended-release drugs), precisely in the topic of formulations with sustained release.\looseness=-1

At the end of the study, the chemist user is able to conclude that metformin is a drug applied in several new formulations platforms, always aiming for its efficient action, prolonged release, or sustained action. Also, this drug is related to patents that intend to its application by non-conventional route of administrations.\looseness=-1

Metformin illustrates the capability of PATopics in the patent group in a correct context, even the patents mentioning the same molecule. Also, this framework promotes important insights about the field of the molecules has been used, innovations regarding the molecule, new applications, new combinations, and the companies involved.\looseness=-1

\subsection{Case study 04 – Comparison between patents}
The PATopics has an extension tool for comparisons between 2 or more patents highlighting their intersection. This application determines the main Topics and subjects involved and to validate this extension, we initially chose two patents to compare. The first patent is -- Sublingual and Buccal Film Compositions, a patent granted in 2019 (belongs to Topic 2), and the second is -- Liquid Naloxone Spray, granted in 2021 (belongs to Topic 19), both patents of the drug Naloxone.\looseness=-1

Figure 5-A shows the comparison and the subjects related to both patents. The two patents converge on Topic 5 and Topic 19, related to patients’ methods and opioids, respectively. Thus, we added a third patent, also about Naloxone -- Pharmaceutical preparation containing oxycodone and naloxone, a patent granted in 2018 (belongs to Topic 13). We observe (Figure 5-B) that the three patents converge only on Topic 19, related to opioids. However, the third patent has an intersection with the first one on Topic 3 (Hypertension, heart conditions, and related conditions) regarding the mention of effects observed in opioid overdose, in which naloxone is the antagonist; and with the second patent, the intersection occurs on Topic 13 (New formulations based in coated or granular vehicles), expected as both are related to new formulations of naloxone.\looseness=-1

\begin{figure*}[htb]
	\centering
	\includegraphics[scale=0.39]{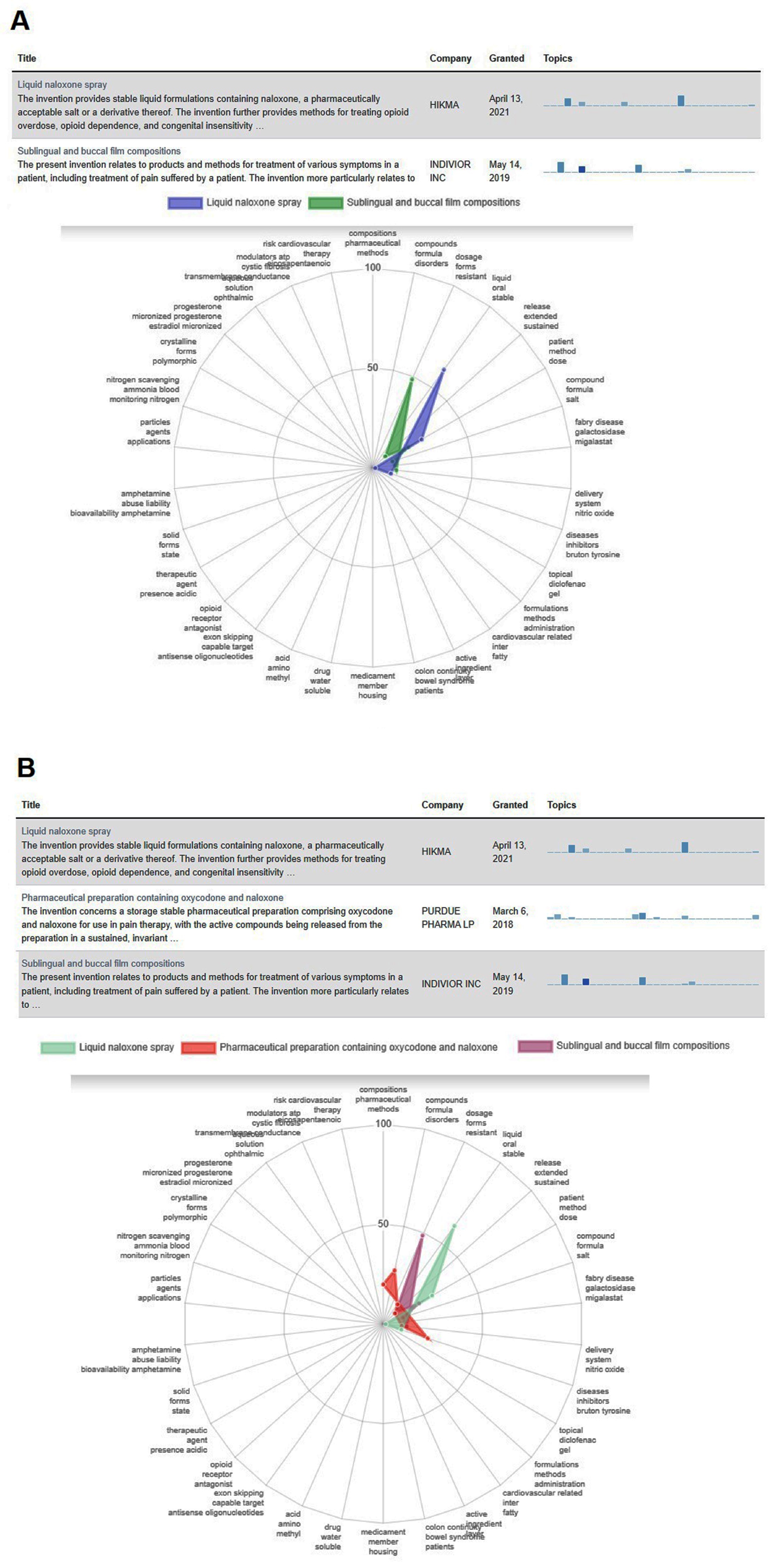}
	\caption{Comparison between patents in which included Naloxone as the drug, we can observe the points in common and differences between the patents regarding the same drug. The first patent is related to a liquid spray, the second is a sublingual and buccal film and the third is a pharmaceutical preparation mixing naloxone with other opioid.}
	\label{fig:5}
\end{figure*}

The comparison tab is an important extension in PATopics of the possibility of details without a complete read of each patent. The patent documents are extensively specific and time-demanding; thus, this application allows the user to get a shortcut in which central points are easily highlighted. We observe how the richest can be the PATopics regarding the discussion, getting insights, and time optimization.\looseness=-1

\section{Conclusion and Future Work}
In this work, we proposed PATopics, a framework specially designed to provide summarized information about pharmaceutical patents. The PATopics framework is inspired by the framework proposed in~\citep{Viegas2022}, and it is composed of four building steps (i) Data representation, (ii) Topic modeling decomposition, (iii) Correlation among entities and (iv) Summary Interface. For data representation, the framework exploits the CluWords representation combined with bi-grams. The CluWords is the state-of-the-art data representation that exploits semantic information provided by word embeddings to enrich the textual information. As for topic modeling decomposition, the framework exploits the Non-negative matrix factorization (NMF), one the most consistent non-probabilistic topic modeling approaches in the literature. The steps Correlation among entities is the specific step that associates the topics retrieved by the NMF method with useful characteristics about the patents, such as inventors, companies, and molecules, while the Summary interface is the friendly web interface that shows all the information for the final user.\looseness=-1

In our experimental evaluation, we instantiated the PATopics Framework using a sample of 4,832 pharmaceutical patents concerning 809 molecules patented by 478 companies extracted from the WizMed platform. We presented how the framework works, showing all the web interfaces of the framework and what information can be extracted from them. In addition, we presented three potential user profiles that can take advantage of PATopics. The first one is the Researcher profile, who works with patents. The second profile is the Chemists, who develop patents, and finally, the Companies or industries who want to use or buy patents. Our analysis showed that PATopics is helpful for all three profiles, especially for researchers and companies, since these two profiles perform more patent searches. Finally, we presented four real-world application cases demonstrating both the practicality and applicability of PATopics in a real scenario. Similarly, this analysis showed that the framework has a good response in meeting the requirements of the use cases.\looseness=-1

For future work, we intend to improve the application by enhancing the topic modeling decomposition step by exploiting novel textual representations~\citep{de2023class,cunha2023tpdr,mendes2020keep}. We intend to build a hierarchical topic of patents~\citep{viegas2020cluhtm}. The hierarchical factor could mitigate the generation of generic topics and bring additional information for correlation entities (third step) since the hierarchical information could be used as an additional component to related patents over distinct layers of the hierarchy topic composition.
\looseness=-1

\section*{Acknowledgments}
This work was partially supported by CNPq, CAPES, Fapemig, AWS, and INWEB.


\bibliographystyle{cas-model2-names}

\bibliography{main}



\bio{figs/bio/pablo_photo}
\textbf{Pablo Cecilio} is a Graduate Student at the Federal University of São João Del Rei, Brazil. He is currently (2023) finishing a bachelor degree in Computer Science.
His main research interests include Machine Learning, Data Mining, Network Systems and UI design.
\vspace{2mm}
\endbio

\bio{figs/bio/juninho_photo}
\textbf{Antônio Pereira} is a Master's student at the Federal University of São João Del Rei, Brazil. Graduated in Computer Science in 2022 and started his master's degree in the same year. His main line of research includes machine learning and data mining with a focus on NLP.
\vspace{2mm}
\endbio

\newpage

\bio{figs/bio/juliana_photo2}
\textbf{Juliana S. Rosa Viegas} is a Post-doctoral Research at Instituto de Investigação e Inovação em Saúde at University of Porto. She hold a Master and Ph.D from School of Pharmaceutical Sciences of Ribeirão Preto, University of São Paulo (Brazil). Her research interests are drug delivery systems specially nanomedicines, their testing in in vitro models, and also biotechnology products.
\endbio

\bio{figs/bio/washington_photo}
\textbf{Washington Cunha} is a Ph.D. Student at the Federal University of Minas Gerais. He got his Master's degree (2019) at the Federal University of Minas Gerais.
His main research interests include Information Retrieval, Machine Learning and Natural Language Processing.
\vspace{2mm}
\endbio

\bio{figs/bio/felipe_photo2}
\textbf{Felipe Viegas} is a Ph.D. student at Federal University of Minas Gerais. He currently works with the design of new data representations for Natural Language Processing applications.  He holds an MS in Computer Science at Federal University of Minas Gerais (2015), and a BS from Federal University of São João Del Rei, Brazil (2012). He has more than eight-year experience in machine learning applications. His research interests include information retrieval and machine learning solutions for textual scenarios.
\endbio


\bio{figs/bio/elisa3}
\textbf{Elisa Tuler} is an associate professor in the Computer Science Department at the Federal University of São João Del Rei, Brazil. She holds a Ph.D. in Information Science from the Federal University of Minas Gerais, Brazil (2015). Her research interests include Data Mining, Human-Computer Interaction and UI design.
\endbio

\bio{figs/bio/fabiana_photo2}
\textbf{Fabiana Testa Moura de Carvalho Vicentini} is an Assistant Professor at School of Pharmaceutical Sciences of Ribeirão Preto, University of São Paulo (Brazil). She received her PhD degree, as well as, developed her postdoctoral research training in Pharmaceutical Sciences at the same University. During this time she did training at Department of Cryobiology and Biomaterials at Institute for Biomedical Tehnologies, Aachen University of Technologie, (Germany) and at Department of Dermatology, University of Michigan (MI, EUA). She coordinated research projects aimed at the production of biopharmaceuticals and worked in the management and quality assurance of biotechnological products and processes. Vicentini research area of interest is focused on Pharmaceutical Nanotechnology with a special approach in Nanovacinology.
\endbio

\bio{figs/bio/leo_photo}
\textbf{Leonardo Rocha} is an associate professor in the Computer Science Department at the Federal University of São João Del Rei, Brazil. He holds a Ph.D. in Computer Science from the Federal University of Minas Gerais, Brazil (2009). His research interests include Information Retrieval, Data Mining, Machine Learning, and Recommender Systems.
\endbio

\end{document}